# Monolithically Integrated Perovskite Semiconductor Lasers on Silicon Photonic Chips by Scalable Top-Down Fabrication

†Piotr J. Cegielski, †Anna Lena Giesecke\*, ‡Stefanie Neutzner, †Caroline Porschatis, ‡Marina Gandini, †Daniel Schall, ‡Carlo A. R. Perini, †Jens Bolten, †Stephan Suckow, §Satender Kataria, †Bartos Chmielak, †Thorsten Wahlbrink, ‡Annamaria Petrozza, †§Max C. Lemme

†AMO GmbH, Otto-Blumenthal-Straße 25, 52074 Aachen, Germany, *E-mail: giesecke@amo.de

‡Center for Nano Science and Technology @Polimi, Istituto Italiano di Tecnologia, via Giovanni Pascoli 70/3, 20133 Milan, Italy

§RWTH Aachen University, Elektrotechnik und Informationstechnik, Lehrstuhl für Elektronische Bauelemente, Otto-Blumenthal-Str. 25, 52074 Aachen, Germany

**Abstract**

Metal-halide perovskites are promising lasing materials for realization of monolithically integrated laser sources, the key components of silicon photonic integrated circuits (PICs). Perovskites can be deposited from solution and require only low temperature processing leading to significant cost reduction and enabling new PIC architectures compared to state-of-the-art lasers realized through costly and inefficient hybrid integration of III-V semiconductors. Until now however, due to the chemical sensitivity of perovskites, no microfabrication process based on optical lithography and therefore on existing semiconductor manufacturing infrastructure has been established. Here, the first methylammonium lead iodide perovskite micro-disc lasers monolithically integrated into silicon nitride PICs by such a top-down process is presented. The lasers show a record low lasing threshold of 4.7 $\mu$Jcm$^2$ at room temperature for monolithically integrated lasers, which are CMOS compatible and can be integrated in the back-end-of-line (BEOL) processes.

**Main text**

Silicon photonics is recognized as a key photonic integration technology due to its compatibility with the CMOS manufacturing infrastructure and the potential for integration with back-end-of-line Si microelectronics, and addresses applications ranging from telecommunications[1] to gas sensing[2] to lab-on-chip.[3] The full potential of this photonic platform is however limited by challenges regarding the integration of laser sources, caused by incompatibilities of active III-V laser materials with silicon technology.[4]

The crystal lattice mismatch between Si and III-V semiconductors leads to defects during epitaxy, which act as nonradiative recombination centers and strongly reduce light emission.[5,6] Despite considerable progress in recent years, the integration of III-Vs by buffered[5] or buffer-less[4] epitaxy cannot meet lattice defect requirements and further exceeds the thermal budget of CMOS electronics. Hence, available systems solutions are limited to III-V wafers bonded to Si waveguide wafers[7] or transfer printed III-V chips.[8]

Metal-halide perovskites in contrast, can be deposited on arbitrary substrates via low temperature (<120 °C), solution based processes, a decisive technological advantage. They are a class of ionic semiconductors with remarkable optoelectronic properties such as carrier diffusion lengths above 1 μm[9] and a direct bandgap,[10] and are investigated for aplications in photovoltaics (PV), photodetectors,[11] LEDs[12] and optically pumped lasers.[13,14] Importantly perovskites also show a promise for electrically pumped lasing, as recently demonstrated with continuous wave (CW) lasing at near-infrared wavelengths below 160 K,[15] CW enhanced amplified spontaneous emission (ASE) in green.[16] and balanced ambipolar transport.[17]

A major bottleneck in perovskite technology is their chemical reactivity, which leads to instability issues and currently prevents manufacturing by well-established semiconductor fabrication processes. In particular, industrial scale manufacturing of perovskite integrated devices would require high throughput, high resolution patterning with precise dimension and overlay control. This is only possible using optical lithography, which includes wet chemical processing in polar solvents and deionized water, which are generally harmful to such highly ionic crystals.[18] Therefore, devices demonstrated so far have been fabricated with alternative techniques such as depositing perovskite on pre-patterned substrates,[19] focused ion beam milling,[20] electron beam lithography using Poly(methyl methacrylate) (PMMA) resist compatible with perovskites,[18] direct imprinting[21] or templated crystal growth.[22]

Here, we present micro disc methylammonium lead iodide (MAPbI$_3$) perovskite lasers manufactured with a conventional top down patterning process including optical lithography. The proposed fabrication scheme enabled the integration of the disc lasers into a silicon nitride photonic waveguide platform. In contrast to previous works on patterned perovskite devices, including our own earlier report, our new technology allows patterning of fully etched structures with high alignment precision and throughput, which is a major technological breakthrough for this technology. The lasing threshold of the fabricated lasers is 4.7 µJ/cm$^{-2}$, which is lower than in unstructured materials.[19,23] Considering that due to processing conditions disc lasers are prone to defects, the obtained results are promising with respect to prospective electrical pumping. Moreover, our fabrication procedure guarantees reproducible results with 83 tested devices fabricated on 4 different PICs exhibiting comparable lasing threshold and spectral characteristics of the lasing modes well controlled with the disc dimensions. In this context, the influence of the thin film polycrystallinity on the device performance was also studied using a model based on randomly generated crystal grains. This simulation approach advances modeling and design of optoelectronic devices based on perovskite thin films.

Experiments were carried out using methylammonium lead iodide (MAPbI$_3$) as the most studied and technologically mature metal-halide perovskite material. Its optical gain is centered at ~785 nm[14]. The design foresees a perovskite micro disc placed in the vicinity of a single mode silicon nitride (Si$_3$N$_4$) bus waveguide, cladded with a planarized SiO$_2$ layer, as shown schematically in **Figure** 1a-b. The disc's edge is aligned with the bus waveguide so that whispering gallery modes (WGM) can be coupled to it by the evanescent field and measured at the waveguide end at a cleaved facet of the photonic chip. The bus waveguide can support both transverse electric (TE) and transverse magnetic (TM) fundamental modes. Resonant modes of the discs (**Figure S 1**, Supporting Information) were simulated using an analytical model[24] and the MAPbI$_3$ refractive index values obtained by ellipsometry (**Figure S 2**, Supporting Information), which are in a good agreement with literature values.[25–27] Calculated free spectral ranges (FSR) between the resonant modes in the targeted wavelength range from 770 to 800 nm were approximately 5.3 nm, 3.3 nm and 2.3 nm for discs with radii of 4 µm, 6 µm and 8 µm, respectively. Precise positions of calculated TE and TM modes, FSR and effective refractive indices ($n_{eff}$) are listed in Tables S 2 and S 3 (Supporting Information). The corresponding $n_{eff}$ of the discs' TE and TM modes were approximately 2.2 and 2.05, while $n_{eff}$ of the single mode bus waveguide modes were only 1.67 and 1.62 for TE and TM polarizations. The lower $n_{eff}$ of the waveguide modes is caused by the lower refractive index of silicon nitride (2.0) with respect to refractive index of MAPbI$_3$ of perovskite (2.56 at 780 nm wavelength, Figure S 2).

The vertical directional coupler was optimized by finite difference time domain (FDTD) simulations via commercially available software (Lumerical FDTD solutions, details in the Supporting Information **Figures S 3-5**). The goal was to optimize the coupling gap and position of the bus waveguide with respect to the disc's edge (called "offset") such that the coupling efficiency was ~1%. The optimum was found for a bus waveguide offset of 150 nm towards the disc center and a separation from the perovskite by a 50 nm gap (Figure 1b). Such positioning precision was not technically achievable with the lithographic tool used in this work, but the disc positions were varied in 250 nm steps over a range of ± 1.25 µm so that there would be at least one nearly optimal structure after fabrication. Coupling efficiencies of both TM and TE WGMs to TM and TE waveguide modes were extracted by the mode expansion method built into the simulation software. The optimized coupling efficiency was 0.57 % for TM and 0.12 % for TE modes. Waveguide modes of opposite polarizations were also excited, albeit weaker by a factor of 6 (TM) and 24 (TE) (referred to as the "cross polarization ratio").

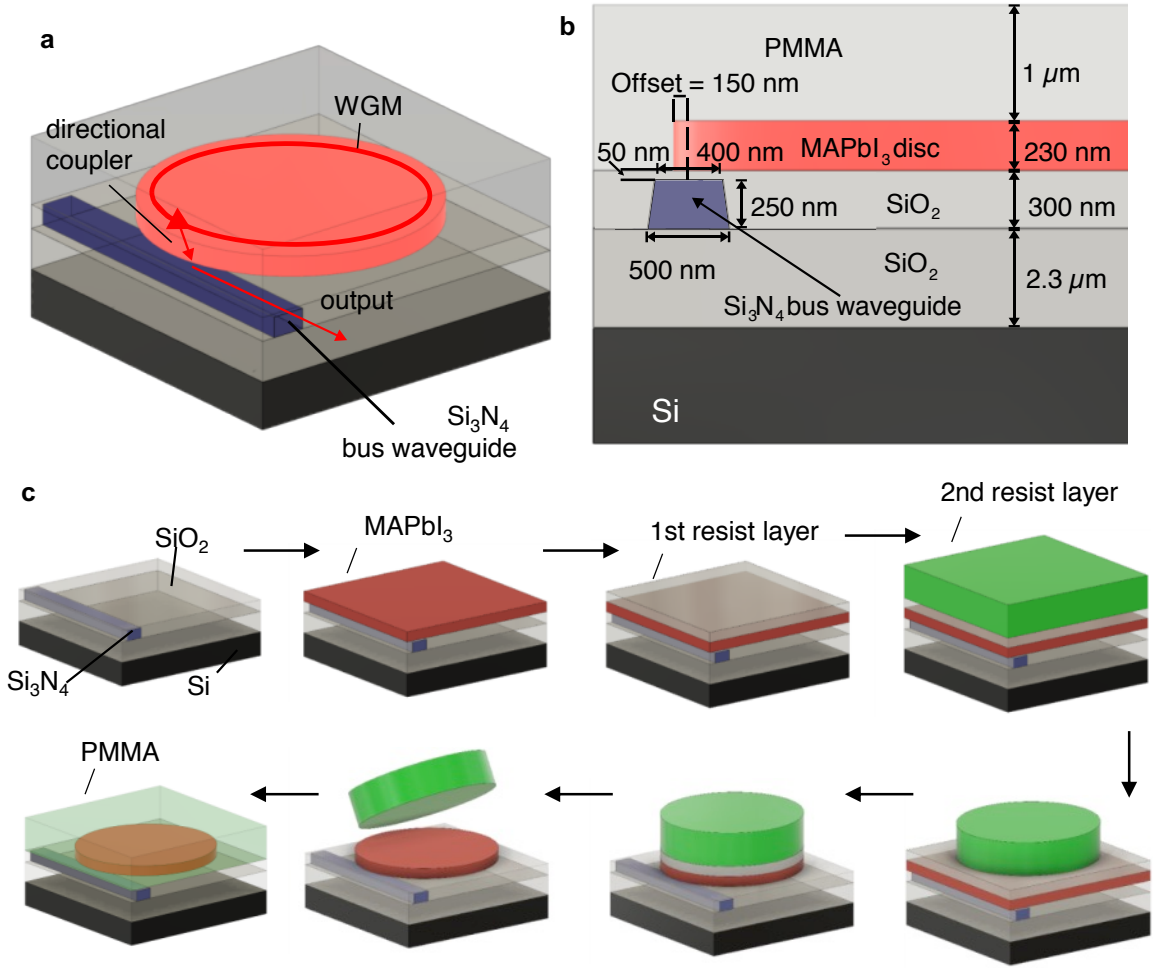

**Figure 1.** a) Sketch of a MAPbI$_3$ disc laser integrated on a silicon nitride photonic chip. b) Cross-section of the disc laser. c) The fabrication process flow (consequent steps marked with arrows): Planarized silicon nitride photonic chip → Deposition of MAPbI$_3$ → Deposition of the bottom resist layer. → Deposition of the top photoactive resist layer. → Developement of the second resist layer. → Pattern transfer by Cl$_2$/CF$_4$ plasma. → Resist stripping by dissolving bottom resist layer and lifting the top layer. → Deposition of 1 $\mu$m thick PMMA cladding

The perovskite disc lasers were fabricated using optical lithography with a double layer resist (details in the Experimental section), where the bottom layer shielded the perovskite from incompatible chemistry used for etch mask definition in the top photoactive layer (Figure 1c). Subsequently the pattern was transferred by a single step ICP RIE process (inductively coupled plasma reactive ion etching) with $Cl_2$ and $CF_4$ gases, which removed the resist interlayer and then the exposed $MAPbI_3$ film. Scanning electron micrographs show that the polycrystalline structure of the initial material is well preserved after patterning (**Figures 2**a-b). The line edge roughness along the edge of the patterned perovskite can be attributed to the roughness introduced by the resist etch mask of the contact lithography process. The perovskite sidewall visible in Figure 2b shows a certain degree of underetching, which increases with $Cl_2$ content in the etch plasma (not shown). The plasma composition was therefore carefully optimized to find a balance between low sidewall damage and the rate of formation of volatile $PbCl_x$ compounds. The samples were further inspected with X-ray diffraction crystallography (XRD) to determine if $MAPbI_3$ degrades during processing (Figure 2c). A small sign of $MAPbI_3$ degradation was detected in the samples after the etching step, recognized by a shoulder in the (110) $MAPbI_3$ peak in the XRD spectrum associated with $PbI_2$. Photoresist stripping did not have an influence on the perovskite. Energy dispersive X-ray (EDX) analysis revealed a trace of chlorine in the micro-discs (Figure 2d, full EDX spectra in **Figure S 6**, Supporting Information).

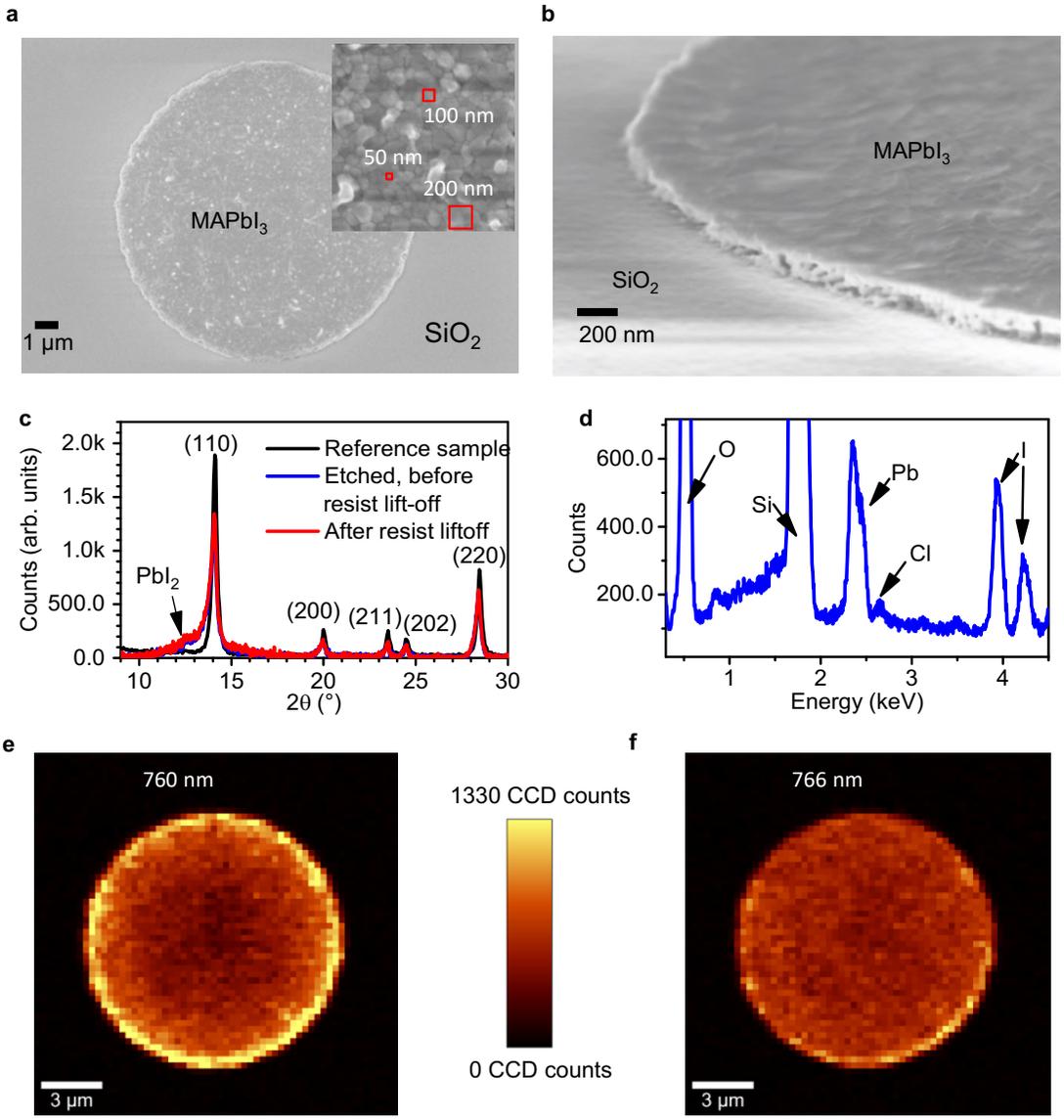

**Figure 2**. Physical characterization of fabricated perovskite micro-discs a) SEM micrograph showing MAPbI₃ disc on SiO₂ substrate after resist stripping. Inset: SEM micrograph at higher magnification depicting perovskite film morphology. Crystal grain size ranged from 50 to 200 nm. b) SEM micrograph of an edge of patterned MAPbI₃ disc after resist stripping taken at 75° angle. c) XRD characterization of a sample containing patterned perovskite structures before (blue) and after (red) the resist stripping compared with a pristine perovskite film (black). d) EDX spectra of the perovskite disc taken close to the discs edge. A trace of chlorine absorbed from the etch plasma

was detected through appearance of a weak peak around 2.62 eV e) Map of photoluminescence of MAPbI$_3$ discs filtered with 15 nm broad filter at 760 nm. c) Map of photoluminescence of MAPbI$_3$ discs filtered with 15 nm broad sigma filter at 766 nm.

The patterned MAPbI$_3$ structures were electro-optically characterized through spatially resolved photoluminescence (PL) spectra (Figure 2e-f, **Figure S 7** and **S 8** of Supporting Information). The PL peak close to the edge of etched MAPbI$_3$ was blue shifted by 6 nm and the PL intensity was twice as high as in the center of the disc. This can be explained by iodine exchange by gaseous chlorine from the etch plasma and formation of MAPbI$_{3-x}$Cl$_x$. Similar effects were reported previously in perovskite films by Solis-Ibarra[28] and in single crystals by Zhang.[29] We have further observed a slight crystal size reduction at the etched edges of the discs, which may also contribute to the blue shift and increase of PL emission.[30]

Disc lasers with radii of 4.3 µm, 6 µm and 8.3 µm, located on separate PICs (among them two PICs containing discs with 6 µm radius), were optically pumped at room temperature in ambient conditions using 120 fs laser pulses at 250 kHz repetition rate centered at 630 nm wavelength (see details in Experimental section). Laser spectra were obtained by collecting light from Si$_3$N$_4$ bus waveguides at the edge of the sample, 5 mm away from the disc, with a single mode polarization maintaining optical fiber featuring a collimator at the other end. The collimated beam entered a spectrometer after passing a polarizer, which was rotated in order to analyze the polarization of light generated in the MAPbI$_3$ discs. Unless otherwise specified, the spectra presented in this work are of TM polarized light (electric field normal to the sample plane).

A broad PL spectrum was observed from the MAPbI$_3$ discs at low excitation fluence (**Figure 3**a, black line). The signal was extremely weak due to inefficient coupling of omnidirectional spontaneous emission into the bus waveguide. When the excitation power was increased, a narrow resonance peak with full width at half maximum (FWHM) of 1.1 nm appeared in the spectrum (Figure 3a, red line). When the emission peak intensity is plotted versus pump pulse fluence, clear thresholds ($P_{th}$) can be extracted (Figure 3b). The discs with radii of 4.3 µm, 6 µm and 8.3 µm show a $P_{th}$ of 7.1 µJcm$^{-2}$ 4.7 µJcm$^{-2}$ and 7.1 µJcm$^{-2}$, respectively. The emission peaks saturate at 18 µJcm$^{-2}$, 11 µJcm$^{-2}$ and 19 µJcm$^{-2}$. Below the lasing threshold we observe a PL signal with a FWHM of more than 20 nm. The spectrum narrows abruptly at a threshold of 4.7 µJcm$^{-2}$, which is accompanied by a superlinear increase of the output intensity (Figure S9, Supporting Information). Both aspects are clear indicators of lasing.[31] The quality factor (Q) of the lasing peak with FWHM of 1.1 nm is 709, which is low for a disc resonator of this size. This is due to the high roughness of perovskite causing significant scattering losses equal to 27.7 dB/cm (Estimated using measured Q factor value, Figure S10 of Supporting Information).

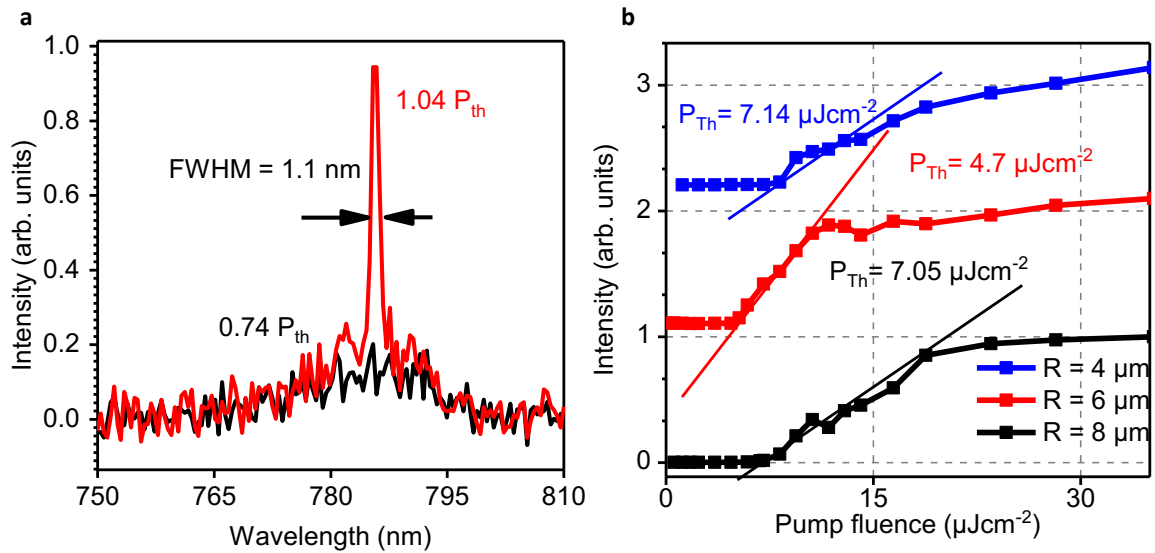

**Figure 3.** a) Emission spectra of a disc with 6 µm radius pumped at 0.74 and 1.04 of the threshold

pump fluence ($P_{th}$). b) Intensity of the emission peak of discs with different radii vs. pump pulse fluence.

In general, the lasing threshold is reached once the modal gain equals propagation and curvature losses. Since the discs with the smallest radius did not have a higher threshold than the larger discs it can be assumed that the curvature loss was not dominant. Hence all lasers would be expected to have similar thresholds since the propagation losses and optical gain depend on morphology, processing and physical properties of the perovskite, which should also be similar. The thresholds of lasers with R = 6 µm (on two different chips) were lower than the thresholds of other lasers. This can be attributed to changing environmental conditions during the deposition process (e.g. varying solvent vapor content in the glovebox atmosphere) influencing the intrinsic materials properties and therefore the optical gain. We also observe a slightly red shifted spectral position of the gain curve compared to the other two samples (**Figure 4**a), which supports this assumption.

At higher excitation of $1.15 \times P_{th}$, additional modes appeared in the spectrum. These allowed investigating the correlation between the free spectral range and the disc radius to identify whispering gallery modes (Figure 4a). The free spectral range was found to be inversely proportional to the disc diameter, following the relation

$$\text{FSR} = \frac{\lambda_1 \lambda_2}{4\pi R} n_g^{-1}, \tag{1}$$

where $n_g$ is the average group index of the modes between two resonant wavelengths $\lambda_1$ and $\lambda_2$ and R is the disc radius (Figure 4b). The measured FSRs are higher than the calculated values. This difference can be attributed to the chlorine intake into $MAPbI_3$ along the etched edge of the structure, where the whispering gallery modes are propagating: the Cl increases the bandgap and results in both lowering of the refractive index[32] and a strong change of the material dispersion at wavelengths considered in the calculations. (Table S 2 and Table S 3, Supporting Information). As FSR is very sensitive to the dispersion (included in Eqn. 1 inside $n_g$ term) the difference between measured and calculated values reached up to 52 %.

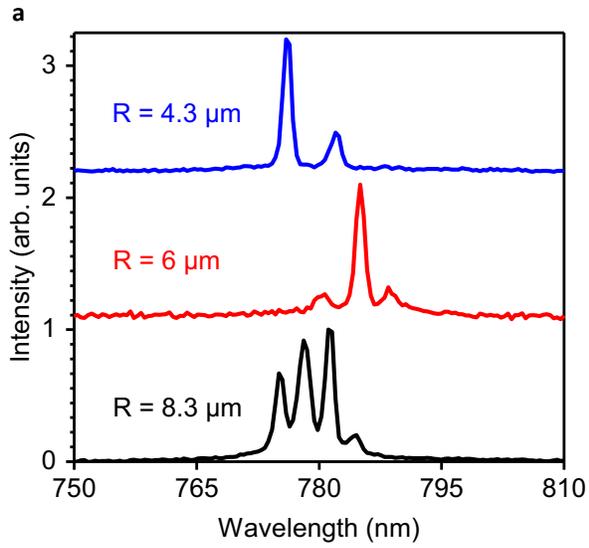

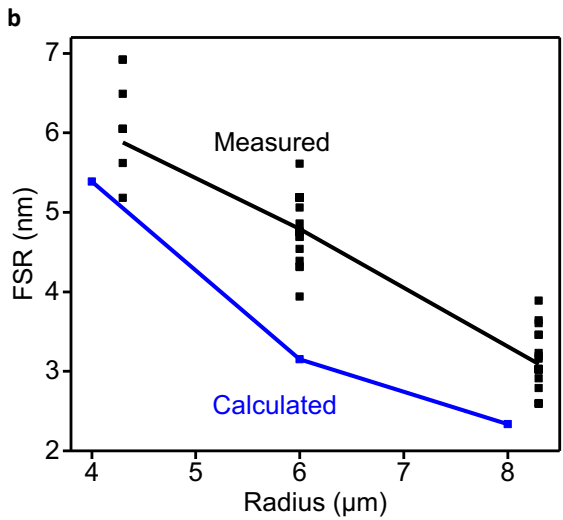

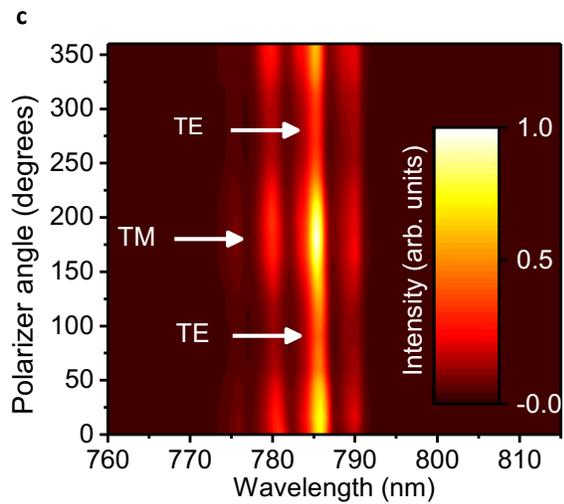

**Figure 4.** a) Spectra of lasers with varying radii located on different samples at excitation of $1.15 \times P_{th}$. b) Average FSR dependency on radius R of the discs (black line) compared with analytically calculated FSR (blue line). Black points show measured FSRs of all 83 measured devices. c) Measured spectra of a disc laser with R = 6 $\mu$m taken at polarizer angle varied with 15° steps.

The laser output was expected to contain two groups of WGMs: TE and TM polarized characterized by different $n_{eff}$ and hence having different sets of resonant wavelengths. Spectra recorded while changing the polarizer angle differed only in intensity of the modes but not in their position in the spectrum (Figure 4c). The measured TM polarized signal was ~3 times stronger than the TE signal, while PL signal below threshold was not polarized (**Figure S 11**, Supporting Information). FDTD simulations serve to explain this behavior, when roughness of the MAPbI$_3$ was included as a parameter (**Figure 5**a). It was implemented by generating random crystal grains of similar size as those observed with SEM in the actual films (Figure 2a-b). The simulated electric field amplitude in the bus waveguide plane was compared for the simple and the polycrystalline case (Figure 5b-c). Despite strong scattering caused by crystal grains, light still couples well to the bus waveguide. This scattering further enables TM and TE WGMs to excite waveguide modes of the opposite polarizations that are 2 and 4.7 times weaker, respectively (**Figure S 12**, Supporting Information ). This unwanted polarization cross-talk is higher than simulated with the simple model (cross-polarization ratios of 2 vs 6.3 and 4.7 vs 23), therefore it can be attributed to the crystal grains. The simulated electric field amplitude profile in the perovskite disc plane (Figure 5d-e) revealed that the crystallites introduce significant optical losses of the WGMs and lead to the excitation of higher order modes. Nonetheless,

WGMs polarization was not altered, indicating that the polarization cross-talk takes place in the directional coupler and not in the disc itself. The roundtrip transmission in a 4.3 μm disc resonator obtained by FDTD simulation with the model incorporating the perovskite's morphology is equal to 0.56, which stands in a good agreement with the round trip transmission of 0.44 extracted from the measurement using the Q factor of 650 measured in a disc of the same size. This confirms the large impact of perovskite roughness on the laser's performance, and that the model correctly resembles the material's morphology.

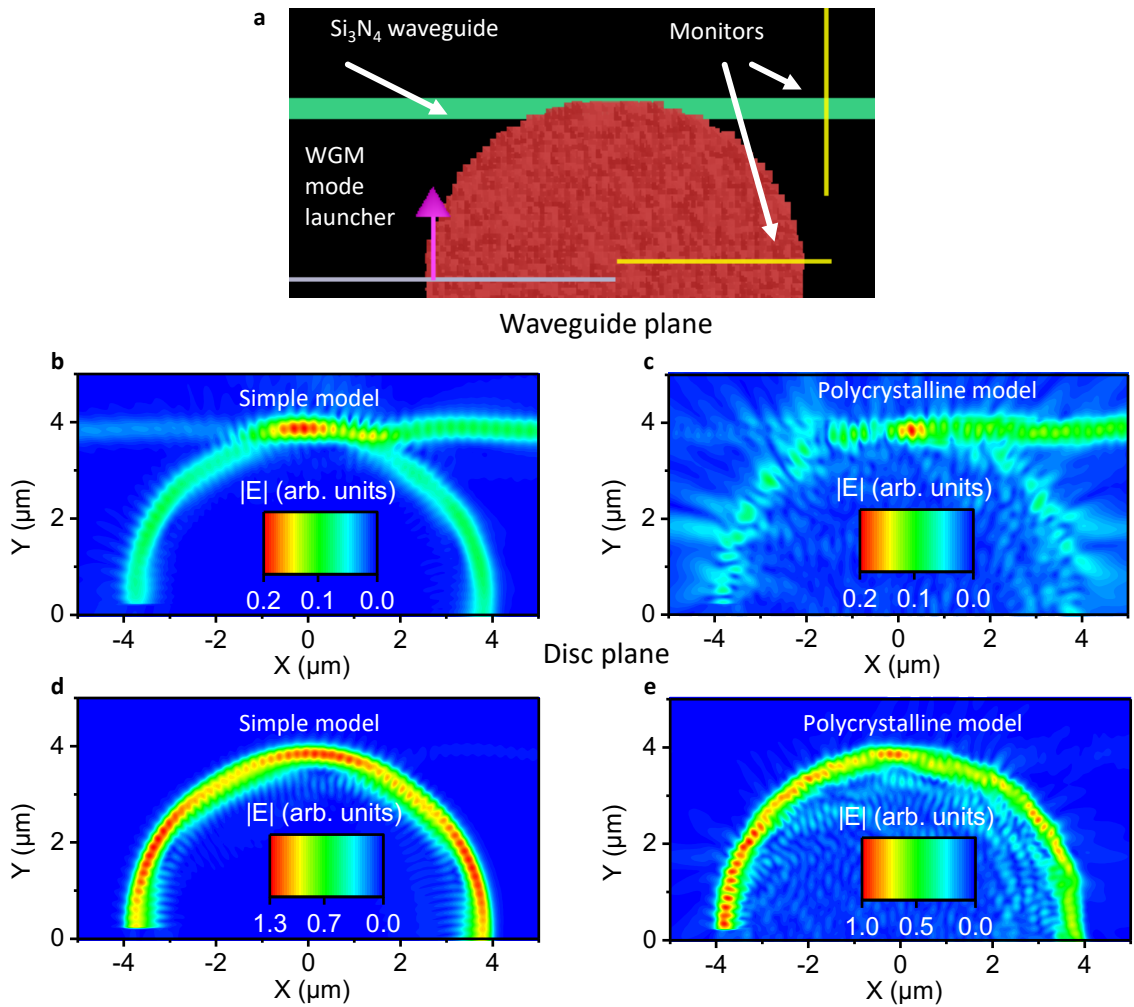

**Figure 5.** a) Model of the disc laser used to simulate directional coupler taking into the account polycrystalline morphology of MAPbI$_3$. b) Electric field amplitude profile in plane of the bus

waveguide obtained by the ideal model. c) Electric field amplitude profile in the bus waveguide plane obtained with polycrystalline model. d) TM WGM in an ideal perovskite disc with no roughness included. e) TM WGM in a polycrystalline disc. Strong scattering and generation of high order modes is visible.

The spectra of the lasers evolved upon increasing the pump fluence, as illustrated in **Figure 6**a. Just above the lasing threshold, there was one clearly dominating mode in the center of the laser spectrum at 785.02 nm. At higher excitation, the mode at 789.51 nm became dominant with a weaker mode appearing at 792.04 nm, at the edge of the optical gain. We attribute this to band gap renormalization, which results in red shifting of the optical gain[33] and can be observed experimentally in the red-shifting of the amplified spontaneous emission (ASE) spectra (**Figure 6**b). One possible explanation of band gap narrowing is heating. However, heating has been shown to widen the band gap in $MAPbI_3$.[34,35] Since the laser operates in high excitation regime at a high electron-hole plasma density, we propose that the band gap narrowing is a result of many-body effects as reported in other direct bandgap semiconductors.[36]

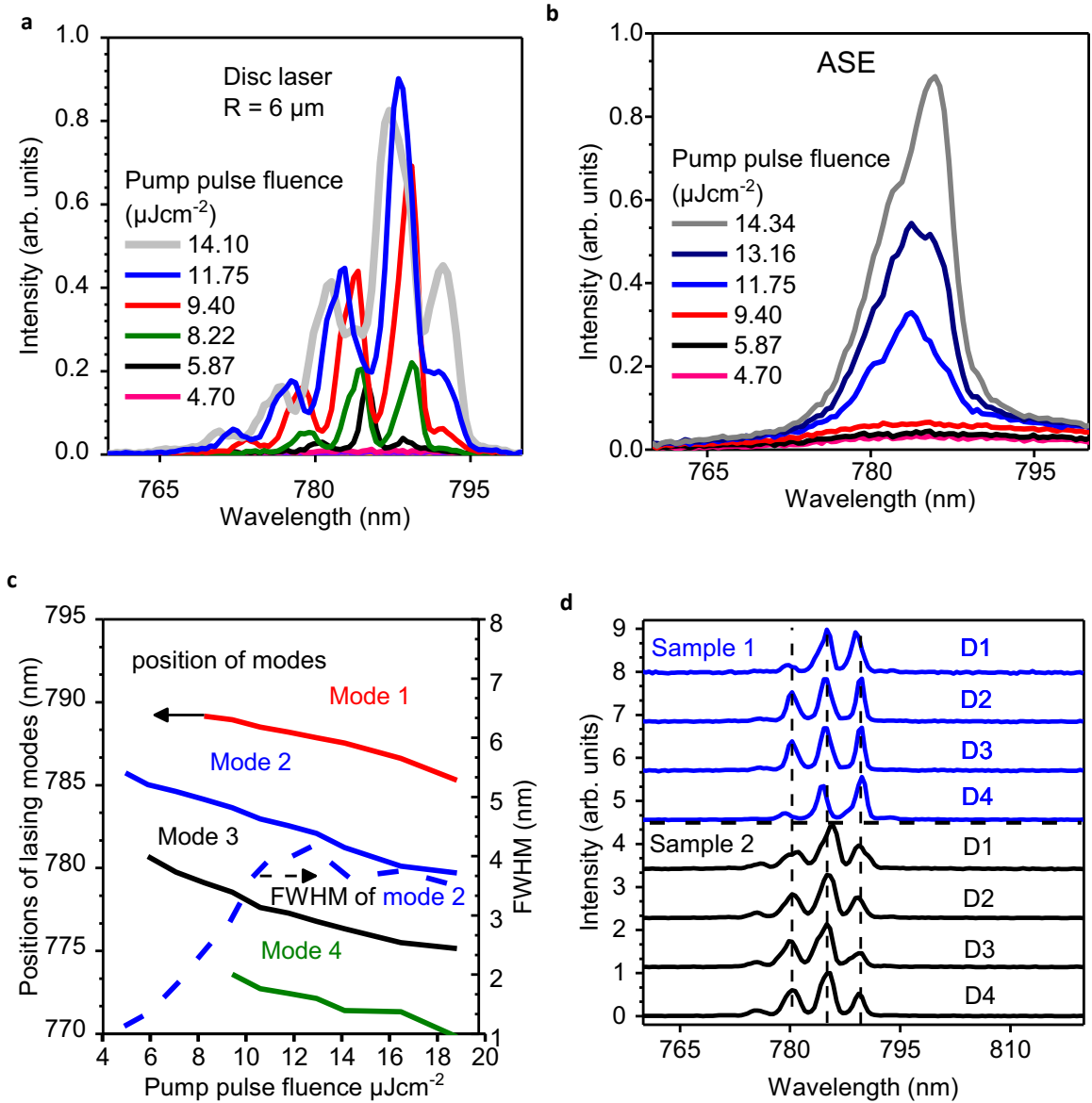

**Figure 6.** a) Emission spectra of a disc laser (R = 6 $\mu$m) compared with b) ASE spectra measured by collecting emission from 15 $\mu$m wide perovskite waveguides. c) Evolution of WGMs with increasing excitation. Solid lines: spectral positions of resonances. Dashed line: FWHM of resonance marked with solid blue line. d) Normalized spectra of lasers with 6 $\mu$m radius located on two different samples. An offset was added for clarity.

The FWHM of the single lasing peak just above the threshold was 1.1 nm and gradually increased to 4 nm at the excitation of ~10 μJcm$^{-2}$ (**Figure 6c**, dashed line), at which point the laser started to saturate and its modes started to overlap (**Figure 6a**, grey line). In a CW laser the FWHM is expected to narrow with increasing pump power due to gain saturation by the lasing mode[37]. In the case of the perovskite disc laser pumped by short pulses, gain does not saturate since the output is multimode. This leads to amplification of spontaneous emission noise which couples to the cavity modes and broadens the laser lines[38]. At very high excitation this ASE noise is so strong that it leads to overlapping of the WGMs (Figure 6a, gray line), in line with the observed increase of FWHM, and observed also in[18]. Furthermore, this parasitic ASE is depleting the available gain and causing the stimulated emission intensity to saturate at an excitation of approximately 11 μJcm$^{-2}$, which coincides with ASE threshold of 10.6 μJcm$^{-2}$ (Figure S13 of the Supporting Information).The spectral positions of the WGMs shifted towards blue with increasing excitation[39] (**Figure 6c**, solid lines) followed by an increase in FSR (e.g. from 5.08 nm at 8.22 μJcm$^{-2}$ to 6.5 nm at 14.1 μJcm$^{-2}$ between modes 2 and 3 ). We attribute this to the change of the dielectric function of the MAPbI$_3$, which modifies the optical length of the circumference of the disc resonators. Above the lasing threshold, in the high excitation regime, the absorption decreases as the material is depleted of carriers, which can absorb photons by excitation from the valence to the conduction band. The refractive index is coupled to the absorption through the Kramers-Kronig relation: the strong reduction of absorption at energies close to the bandgap where the laser is operating induces a significant reduction of the real part of the refractive index,[40] causing blue shifting of the resonant modes. In summary, the spectral characteristics of the micro disc emissions include spectral narrowing, polarization and verified WGMs, which - in conjunction with a clear threshold - confirms lasing in our devices.

The overall reproducibility achieved with the technology presented in this work is an outcome of many optimized processing steps, such as fabrication of the photonic chips, perovskite deposition and perovskite patterning. A comparison of discs with R = 6 µm located on two different samples confirms the reproducibility of the MAPbI$_3$ patterning (**Figure 6**d). The largest source of variability was the polycrystalline morphology of MAPbI$_3$. The crystal size distribution ranged from 50 nm to 200 nm (**Figure 2** a-b), which translates into fluctuations of the refractive index within disc resonators and non-uniformities of the discs thicknesses. This resulted in variations of the group index of WGMs, which became manifest in a standard deviation of FSR reaching 1.3 nm within a single sample. The differences in MAPbI$_3$ morphology between different samples led even to different spectral positions of gain,[41] as observed in **Figure 4**a. We demonstrated a scalable, high throughput technology for patterning of metal-halide perovskites with industry standard processes. The methodology was applied to lead halide perovskites that were patterned into MAPbI$_3$ disc lasers. These lasers were integrated into a silicon nitride waveguide platform, where they were characterized to benchmark the technology. We achieved a lasing threshold of 4.7 µJcm$^{-2}$ and a linewidth of 1.1 nm, which is one of the lowest reported values for polycrystalline perovskites lower than in devices made of unprocessed perovskites[19,23] and is the record value for monolithicaly integrated CMOS compatible lasers.[42] High device-to-device and sample-to-sample reproducibility evidences that this technology has potential for industrial scale fabrication. We further introduce an optical simulation method which takes into account the polycrystalline nature of solution processed perovskite thin films and thus supports understanding the behavior of experimental devices. The tool can be further exploited to design perovskite electro-optic components based on polycrystalline thin films.

This work provides a large step towards exploiting the main advantages of organo-metal perovskites over III-V semiconductors, e.g. low cost deposition from solution and straightforward bandgap engineering, to provide cost efficient lasers at infra-red and visible wavelengths.[43] The disk lasers investigated here demonstrate an important technological advancement, i.e. the feasibility of process integration of perovskite materials with high throughput patterning techniques. This work thus removes a major obstacle on the path towards exploitation of these materials in commercial integrated optoelectronic circuits and points towards developing techniques for wafer scale deposition of high quality perovskite films. Furthermore, our technology can be used as a base for development of electrically pumped lasers. This will include not only the development of efficient carrier injection into the perovskite but also , following the footsteps of the perovskite solar cell community[44], detailed studies of stability and degradation mechanisms of perovskites as laser gain media. Considering requirements for reliability of commercially available laser diodes with wear out times of up to $10^6$ hours[45], preventing degradation by implementation of hermetic packaging[46] preventing moisture associated deterioration[47] and by barriers stopping corrosion of metal electrodes by e.g. iodine ions[48] is of great importance for perovskite lasers. Our lasers, protected only by a 1 μm thick PMMA coating, were intensively tested in ambient conditions for multiple hours and stored in a nitrogen filled box in between the experiments. These were carried out daily for three weeks without significant performance deterioration or a single device failing. This is a promising result indicating that reaching reliability required in commercial applications should be possible in the future.

**Experimental section**

*Fabrication of PICs:* Devices were fabricated on 6" silicon wafers with 2.2 μm of thermally grown $SiO_2$. 250 nm of silicon nitride was deposited by a low pressure chemical vapor deposition (LPCVD) process from dichlorosilane and ammonia gases. The $Si_3N_4$ waveguides were defined by AZ-MiR 701 photoresist and reactive ion etched with an inductively coupled plasma (ICP) using $CFH_3$ and helium gases. This was followed by a LPCVD deposition of 600 nm of $SiO_2$ (low temperature oxide - LTO) by silane and oxygen reaction. Planarization of the $SiO_2$ surface was done by spin coating of spin-on glass (SOG) and subsequent etching in $CHF_3$ plasma. This process cycle of SOG coating and dry etching was repeated twice, resulting in approximately 50 nm height difference between the LTO surface above and next to the silicon nitride waveguides with a smooth transition. Hence, about 300 nm of LTO was consumed during these planarization steps leaving 50 nm of LTO above silicon nitride waveguides. Next, wafers were coated with protective polymer and diced into 3×3 cm samples.

*Perovskite deposition*: The photonic chips were cleaned from the protective polymer layer through sonication in acetone and isopropanol, followed by 10 minutes of oxygen plasma cleaning in a RF plasma reactor. The chips were then brought into a nitrogen filled glovebox for the subsequent perovskite deposition. The perovskite solution was prepared with 1M of methylammonium ($CH_3NH_3I$, Dyesol), lead iodide ($PbI_2$, 99.9985 %, Alfa-Aesar) and dimethyl sulfoxide (DMSO, anhydrous, ≥99.9 %). These were dissolved in 1ml of dimethylformamide (DMF, anhydrous, 99.8 %) and heated to 60 °C for 1 hour while stirring. The resulting solution was filtered (0.22 um polytetrafluoroethylene (PTFE) filter) and spin-coated on the cleaned silicon substrates at 6000 rpm for 30 s. 300 μl of chlorobenzene (anhydrous, 99.8 %) was dripped after 6 s from the spin start. The resulting film was exposed to a two-step annealing process (45 °C, 5 min + 100 °C, 10 min) to complete the conversion of the precursors into perovskite.

*Patterning of $MAPbI_3$:* First, a 200 nm resist layer was deposited via spin coating. AZ MiR 701 positive tone photoresist was then spin coated at 3000 rpm and annealed at 95 °C for 90 s. Next, the micro disc lasers were defined through a mask by UV light using a contact lithography mask aligner (EVG 420). Afterwards the photoresist was baked at 115 °C for 90 s and developed for 20 s in MF26 CD developer. The pattern was transferred into the PMMA/$MAPbI_3$ stack in a single reactive ion etch step with $Cl_2$ and $CF_4$ gases. The chlorine forms volatile $PbCl_x$ compounds[49] while $CF_4$ is used to bind and neutralize excess chlorine to prevent under etching. The photoresist was removed by dipping the samples in a non-polar solvent at 40 °C for 20 minutes. Immediately after removing samples from chlorobenzene, 1 μm of PMMA was spin coated and annealed at 80 °C for 10 minutes.

*Photoluminescence mapping*: Photoluminescence mapping was performed using a WiTec Confocal Raman microscope. The sample was scanned with a 532 nm CW laser at 100 nW power. The laser spot was 300 nm in diameter and the scanning step size was 400 nm. The measurement was taken at room temperature in ambient conditions.

*XRD, EDX and SEM inspection*: XRD spectra were taken on complete devices before and after the etching process using a Bruker D8 Advance (Bragg-Brentano geometry). The EDX measurements were performed in an SEM Zeiss SUPRA 60 with a Bruker XFlash 4030 detector. The primary energy was 15 keV, the takeoff angle was 35° and the working distance was 12 mm. SEM micrographs were taken using the same microscope at 4 kV and a working distance of 3.5 mm.

*Laser characterization:* The disc lasers were pumped with 120 fs pulses at 630 nm wavelength and repetition rate of 250 kHz delivered by a commercial optical parametric amplifier (OPA, Coherent 9450). The OPA was provided with 50 fs pulses at 800 nm wavelength generated by a mode-locked Ti:sapphire oscillator (Coherent Micra) in conjunction with a regenerative amplifier (Coherent RegA 900). The pump laser beam was focused with a 150 mm lens before the sample. The samples were positioned on a sample chuck mounted on a XY micrometer stage and held by vacuum. To guarantee safe alignment of the optical fiber end to the sample edge, a CCD camera with 8× zoom lens with a working distance of 82 mm was used. We calculated the Gaussian spot size of the elliptically shaped laser spot to be 9860 $\mu m^2$, using an image of photoluminescence from the excited perovskite. The output was collected by edge coupling to a polarization maintaining fiber (Thorlabs PM780-HP, mode field diameter 4.9 μm) and collimated by Toptica Fiberout_01227. The collimated beam passed through a metal wire grid polarizer and was recorded by Maya2000 Pro (Ocean optics) spectrometer with 0.4 nm resolution.

All samples were measured in air at room temperature.

## ASSOCIATED CONTENT

Additional information on details of calculations, simulations material and device characterization (PDF).

## AUTHOR INFORMATION

### Corresponding Author

*Dr. Anna Lena Giesecke, AMO GmbH, Otto-Blumenthal-Str. 25, 52074 Aachen, Germany

### Author Contributions

P.J.C proposed the device and the patterning process, performed simulations and designed the lasers. P.J.C., C.P and B.C. fabricated silicon nitride photonic chips. M.G. and C.A.R.P. optimized and produced the perovskite layers. P.J.C. and D.S. optimized the perovskite patterning process. S.K., J.B., C.A.R.P. and P.J.C. performed the material characterization. P.J.C. and S.N. characterized the micro disc lasers. P.J.C., A.L.G., and S.S. analyzed the measurement results. A.L.G., A.P., T.W. and M.C.L coordinated the work. All authors discussed the results and contribute to writing of the manuscript.

## ACKNOWLEDGMENT

Funding by the European Union's Horizon 2020 research and innovation programme under the Marie Sklodowska-Curie grant agreement No 643238 (Synchronics) and PPP No 688166 (Plasmofab) is gratefully acknowledged.

Table of contents graphic:

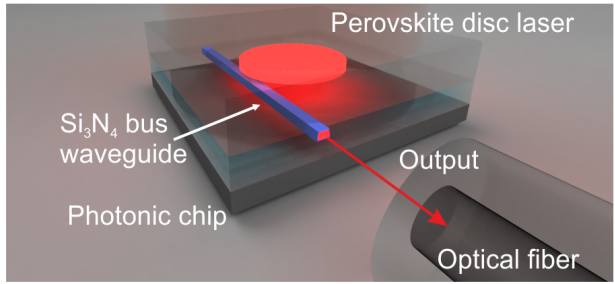